\documentclass[journal]{IEEEtran}

\usepackage[english]{babel}
\usepackage{amssymb}
\usepackage{amsmath}
\usepackage{amsthm}
\usepackage{mathrsfs}
\usepackage{epsfig}
\usepackage{graphicx}
\usepackage[thinlines]{easytable}
\usepackage{enumitem}
\usepackage{accents}
\usepackage{textcomp}

\newlength{\dhatheight}

\graphicspath{{./figs/}}

\newcommand{\eg}{e.g.,\ }
\newcommand{\ie}{i.e.,\ }

\title{Wireless Multi-Sensor Networks for Smart Cities:\\ A Prototype System with Statistical Data Analysis} % Title, preferably not more

\author{Bal\'azs Cs. Cs\'aji, \IEEEmembership{Member,~IEEE},\hspace{5mm} \and Zsolt Kem\'eny,\hspace{5mm} \and Gianfranco Pedone,\hspace{5mm} \and Andr\'as Kuti, \hspace{5mm} \and J\'ozsef V\'ancza\vspace{-2mm}%
\thanks{The work has been supported by the National Research Development and Innovation Office under contract number ERNY\H{O}\_13\_1\_2013-0011. Authors from MTA SZTAKI also thank for the support of the GINOP-2.3.2-15-2016-00002 grant. B.~Cs.~Cs\'aji was supported by the J\'anos Bolyai Research Fellowship of the Hungarian Academy of Sciences, BO\,/\,00217\,/\,16\,/\,6.}
\thanks{Bal\'azs Cs. Cs\'aji, Zsolt Kem\'eny, Gianfranco Pedone, and J\'ozsef V\'ancza are with MTA SZTAKI: The Institute for Computer Science and Control of the Hungarian Academy of Sciences, Kende utca 13-17,
1111 Budapest, Hungary; emails: \texttt{\{balazs.csaji, zsolt.kelemy, gianfranco.pedone, jozsef.vancza\}@sztaki.mta.hu}}%
\thanks{Andr\'as Kuti is on leave from General Electric (GE) Hungary Ltd., V\'aci \'ut 77., 1044 Budapest, Hungary; email: \texttt{kutiandras@gmail.com}}}

\begin{document}
\maketitle

%%%%%%%%%%%%%%%%%%%%%%%%%%%%%%%%%%%%%%%%%%%%%%%%%%%%%%%%%%%%%%%%%%%%%%%%%%%%%%%%%%%%%%%%%%%%%%%%%%%%%%%%%%%%%%%%%%%%%%%%%%%%%%%%%%%

\begin{abstract}
As urbanization proceeds at an astonishing rate, cities have to continuously improve their solutions that affect the safety, health and overall wellbeing of their residents. Smart city projects worldwide build on advanced sensor, information and communication technologies to help dealing with issues like air pollution, waste management, traffic optimization, and energy efficiency. The paper reports about the prototype of a smart city initiative in Budapest which applies various sensors installed on the public lighting system and a cloud-based analytical module.

While the installed wireless multi-sensor network gathers information about a number of stressors, the module integrates and statistically processes the data. The module can handle inconsistent, missing and noisy data and can extrapolate the measurements in time and space, namely, it can create short-term forecasts and smoothed maps, both accompanied by reliability estimates. The resulting database uses geometric representations and can serve as an information centre for public services. 

\end{abstract}

%%%%%%%%%%%%%%%%%%%%%%%%%%%%%%%%%%%%%%%%%%%%%%%%%%%%%%%%%%%%%%%%%%%%%%%%%%%%%%%%%%%%%%%%%%%%%%%%%%%%%%%%%%%%%%%%%%%%%%%%%%%%%%%%%%%


\IEEEpeerreviewmaketitle

%%%%%%%%%%%%%%%%%%%%%%%%%%%%%%%%%%%%%%%%%%%%%%%%%%%%%%%%%%%%%%%%%%%%%%%%%%%%%%%%%%%%%%%%%%%%%%%%%%%%%%%%%%%%%%%%%%%%%%%%%%%%%%%%%%%
%%%%%%%%%%%%%%%%%%%%%%%%%%%%%%%%%%%%%%%%%%%%%%%%%%%%%%%%%%%%%%%%%%%%%%%%%%%%%%%%%%%%%%%%%%%%%%%%%%%%%%%%%%%%%%%%%%%%%%%%%%%%%%%%%%%

\section{Introduction}
\IEEEPARstart{A}{smart city} 
is an urban environment which
combines
advanced 
sensor, information and communication technologies to help efficiently manage the assets of the city. These include {\em services} related to 
health, transportation, sustainability, economy, law enforcement, community and others affecting the overall wellbeing of the residents and businesses  \cite{Tryfonas:2014,Schneider:2014}.

Sensor networks are crucial components of smart cities as 
the data they gather 
are fundamental for these services.
 {\em Wireless sensor networks} (WSNs) are especially important as they can be built by relatively cheap and small sensors with low power consumption and 
maintenance cost whose ability to transmit data remotely allows their  deployment at a large variety of locations. Some applications of WSNs in smart cities include pollution prevention, waste management, structural health monitoring, smart buildings, 
surveillance,
intelligent transportation,
traffic light control, parking optimization, environmental monitoring, 
and energy management \cite{oliveira2011wireless,hancke2012role}.

As data can come from various sources, building systems which can {\em integrate} data of diverse origin, e.g., measurements from a multi-sensor network, are of increasing interest \cite{puiu2016citypulse}.

There are several {\em intelligent lighting} projects where various sensors are installed on the public lighting system \cite{shahzad2016energy,kovacs2016intelligent,lavric2014street,mahoor2017hierarchical} or in smart buildings \cite{viani2017evolutionary}, in order to improve the efficiency of the lighting service. They typically integrate luminaries using solid state light emitting diode (LED) technology as it allows smart dimming control \cite{kovacs2016intelligent}. These systems primarily apply sensors to measure environmental light, power consumption, and the presence of traffic or people, based on which they can control the system in order to increase the quality and the energy-efficiency of the service \cite{viani2017evolutionary,viani2016experimental}. 

Another key area of WSNs for smart cities is {\em environmental monitoring} both outdoor and indoor. In traditional applications it is done by a small number of expensive, high-precision sensors; while WSNs offer a promising alternative by using a large number of low-cost, average-precision units instead \cite{oliveira2011wireless}.

In this paper a prototype smart city initiative is presented in which a wireless multi-sensor network is installed on the public lighting system in Budapest, Hungary, for environmental monitoring. The system also includes an analytical module which performs statistical data analysis on the gathered data. 

The {\em key features}
of the presented prototype are as follows:
\smallskip
\begin{itemize}
\item The installed WSN includes a wide range of sensors measuring many {\em air quality} and {\em traffic} related stressors. This allows the simultaneous {\em monitoring} and potential analysis of various phenomena, including their inter-dependencies and their (joint) temporal and spatial dynamics. \smallskip
\item The cloud-based {\em analytical module} periodically analyses the data gathered by the WSN (it currently focuses on air quality related stressors). The module generates efficient short-term {\em forecasts} and smoothed {\em maps}, both accompanied by {\em reliability} estimates; and stores the results in a dedicated database using {\em geometric} representations which allows {\em flexible queries}. The module can also deal with {\em inconsistent}, {\em missing} and {\em noisy} measurements.
\end{itemize}
\smallskip

Hence, instead of gathering data for a specific pre-defined application, the presented prototype measures a broad variety of quantities leaving open the potential services that they can support. The main novelty of the system is the cloud-based 
data processing 
unit that 
statistically analyses the measurements and makes its results available in a spatial database using flexible representations, such as polyhedral surfaces and line strings. This simplifies the application of the data and makes it more attractive for potential public services.

The paper is organized as follows. First, the architecture, underlying infrastructure and quantitative considerations are summarized. Next, pre-processing, modelling and forecasting techniques are presented.
Then, 
map generation methods are discussed, while the paper ends with experimental validations.

%%%%%%%%%%%%%%%%%%%%%%%%%%%%%%%%%%%%%%%%%%%%%%%%%%%%%%%%%%%%%%%%%%%%%%%%%%%%%%%%%%%%%%%%%%%%%%%%%%%%%%%%%%%%%%%%%%%%%%%%%%%%%%%%%%%

\section{General Concept}
The results 
of this paper were obtained with a pilot implementation of a WSN deployed in a real urban environment, relying on commercial wireless network and electric lighting infrastructure, yet, being an {\em experimental prototype} with regard to sensor coverage and implementation of information processing solutions. Aside from tests with system performance and choice of measurement hardware, certain types of monitoring and analytical services,
e.g., short-term forecasts, smoothed maps, reliability estimates, were in the focus of the research. 

Higher-level end-user services, such as decision support, early warnings, or access interfaces for the municipality and individual citizens, were excluded,
but later they could be provided as services based on the maintained analytical database.

\subsection{Hardware and Network Architecture}
\label{subsec:hardware}

About 700 sensors at 70 locations have been installed on the public lighting system in District XII of Budapest, Hungary; with a plan to extend the number of locations to 250. The sensor boxes were mounted on light poles about 8.5 meters off the ground. The sensors cover approximately one square kilometer around a shopping mall including six main streets and two squares. The distance between neighboring sensors is typically between 50 and 200 meters. The first sensors have been installed during the summer of 2015, while the data which the presented research is based on were mainly gathered from January, 2016 till early November, 2016 (ten months).

\begin{figure}[b]
 \vspace{-1mm}
 \centering
 \includegraphics[width=1\linewidth]{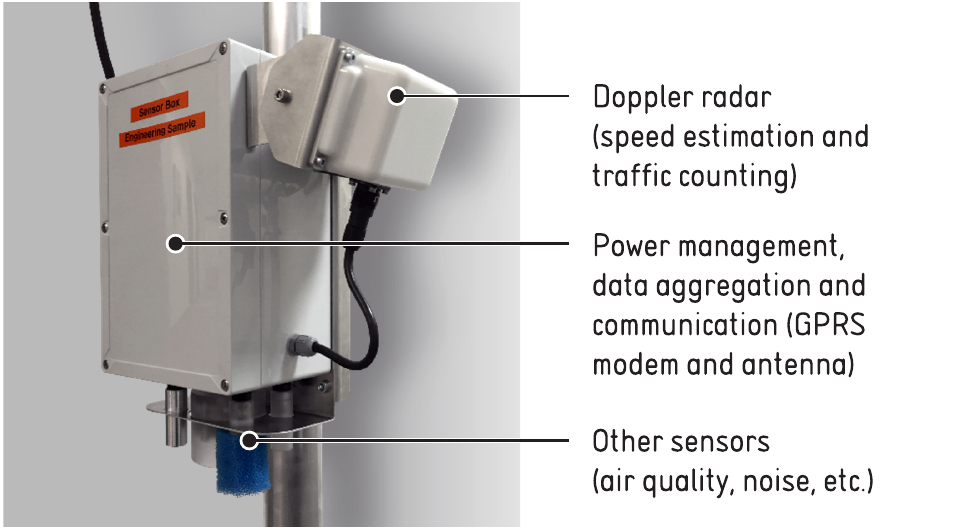}
 \caption{Sensor box used in the project installed on a light pole.}
 \label{fig:sensorbox}
 \end{figure}

The WSN delivering measurement data for the project was designed with the objective of using existing sensor accommodation, power and communication resources as far as circumstances allow. 
Therefore, the individual sensors were installed in low-maintenance sensor boxes with sufficient local computing power to bundle and transmit measurement data, see Figure \ref{fig:sensorbox}. The sensor boxes are installed on selected luminaries of the public lighting system, with access to the power grid while the street lights are in use. In addition, the sensor boxes are also provided with their own batteries and power management system, allowing independent operation during daytime when the luminaries are powered off.

Table \ref{tab:stressors} overviews the measured {\em stressors} with their sampling intervals. The first seven are air quality while the last three are traffic related. Vibratory acceleration (of the sensor box) and speed (of the vehicles passing by) are measured in all (x, y, z) directions, furthermore, the speed and noise senors provide histograms about the measured quanties in their sampling intervals that can be used, e.g., for traffic counting.

\begin{figure}[t]
\begin{center}
\includegraphics[width=\columnwidth]{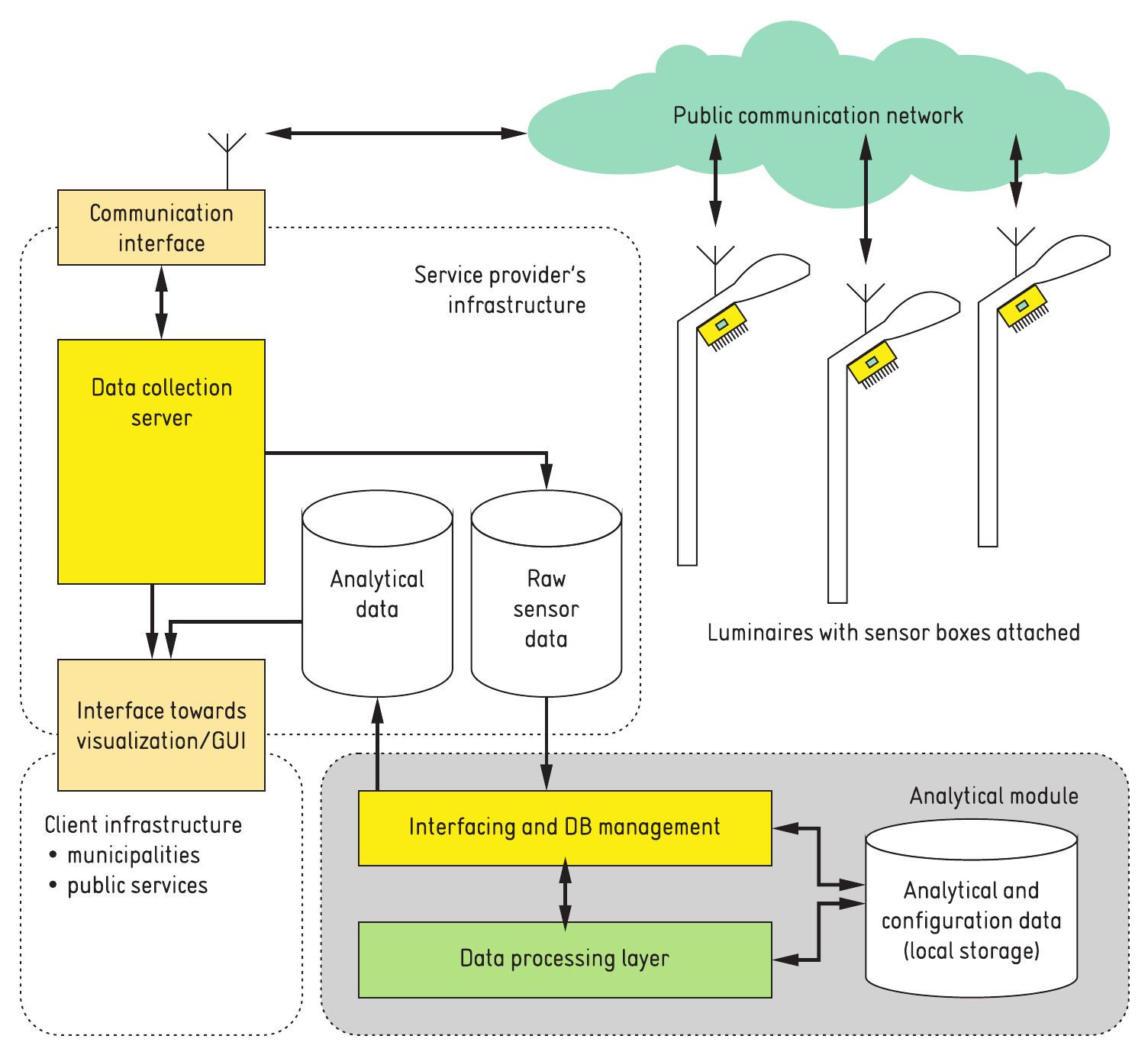}
\caption{General overview of the prorotype system including the wireless multi-sensor network, the databases and the analytical module.}
\label{fig:netarch_experimental}
\vspace*{-3mm}
\end{center}
\end{figure}

The WSN has a {\em star topology}, namely, each unit has a direct connection to the data center. Transmission of measurements occurs individually for each sensor box, using a public mobile 
network at low priority (\ie data may be lost when the network is at peak load). Transmitted data are received by the {\em data collection server} which stores the type of physical\,/\,statistical quantity, the measured\,/\,calculated value, and the time stamp of the measurement in a dedicated database, see Figure \ref{fig:netarch_experimental}, providing the input for analytical functionalities.

{\renewcommand{\arraystretch}{1.2}
	\begin{center}
		\begin{table}[b]
			\vspace*{-3mm}
			\caption{Summary of the measured stressors }
			\centering
			\begin{tabular}{|l||c|c|c|}\hline
				& Default &\multicolumn{2}{c|}{Sampling interval}\\ \hline			
				Stressor & unit & \hspace{0.5mm}minimum \hspace{0.5mm}& \hspace{0.5mm}maximum \hspace{0.5mm} \\ \hline\hline
				particulate matter      & g\,/\,m\textsuperscript{3} & 10 min & 60 min  \\ \hline
				environmental temperature  & \textdegree{}C & 1 min & 5 min   \\ \hline
				ultraviolet irradiation (B) & index  & 10 min & 30 min   \\ \hline
				ambient light             & lux  & 10 min  & 30 min  \\ \hline
				air pressure              & mbar  & 1 min  &  5 min   \\ \hline
				relative humidity       &  \% & 1 min  & 5 min    \\ \hline
				carbon monoxide      & ppm &  30 min &  60 min \\ \hline\hline
			    noise (histogram)      & dBA  & 1 min  & 15 min   \\ \hline	
				speed (x, y, z; histogram)      & km\,/\,h  & 1 min  & 15 min  \\ \hline
			   	vibratory acceleration (x, y, z) & mG & 1 min  & 10 min  \\ \hline
			\end{tabular}
			\label{tab:stressors}
		\end{table}
\end{center}}

\vspace{-8mm}
\subsection{Analytical Module}
The sensor boxes are responsible for the data collection, initial processing and aggregation, and client communication; on the other hand, the deeper analysis of the data is provided by a software module located in a cloud infrastructure.

Therefore, the analytical functionalities are accommodated in a dedicated \emph{analytical module}, cleanly separating the resources of the service provider and the experimental setup. Configuration data related to sensors and periodically invoked algorithms, exclusively used by the module, are stored locally, while the results of short-term forecasts and extrapolation on a grid of geographical locations, both accompanied by reliability estimates, are returned to the service provider, and are stored 
in an analytical 
database, see Figure \ref{fig:netarch_experimental}. 
In order to facilitate the efficient representation as well as flexible and complex queries of spatio-temporal data, special encodings involving geometrical objects are used, see Section \ref{subsec:gis}.

The analytical module is interfacing with the service provider's infrastructure through the aforementioned 
databases only, and carries out \emph{batch processing} steps invoked at scheduled points of time. The main steps of the module are
\medskip
\begin{enumerate}
\item loading the {\em configuration} table from the database;
\item reading out the {\em measurements} from the input database;
\item {\em pre-processing} the observations (including discretization, denoising, and estimating missing values);
\item fitting (discrete-time) stochastic {\em time-series} model(s);
\item computing (short-term) {\em forecasts} with confidence regions;
\item creating extrapolated {\em maps} with reliability estimates;
\item {\em post-processing} the generated forecasts and maps;
\item storing the results in the output {\em database} (using GIS).
\end{enumerate}
\medskip
The current implementation of the analytical module focuses on analysing stressors related to {\em air quality}
(i.e., the first seven quantities of Table \ref{tab:stressors}, 
cf.\ Tables \ref{tab:rmse} and \ref{tab:map-rmse}), while providing traffic specific analyses is a potential future work.

%%%%%%%%%%%%%%%%%%%%%%%%%%%%%%%%%%%%%%%%%%%%%%%%%%%%%%%%%%%%%%%%%%%%%%%%%%%%%%%%%%%%%%%%%%%%%%%%%%%%%%%%%%%%%%%%%%%%%%%%%%%%%%%%%%%

\section{Cloud-Based Computational Platform}
\begin{figure}[t]
\centering
\includegraphics[width=1.0\linewidth]{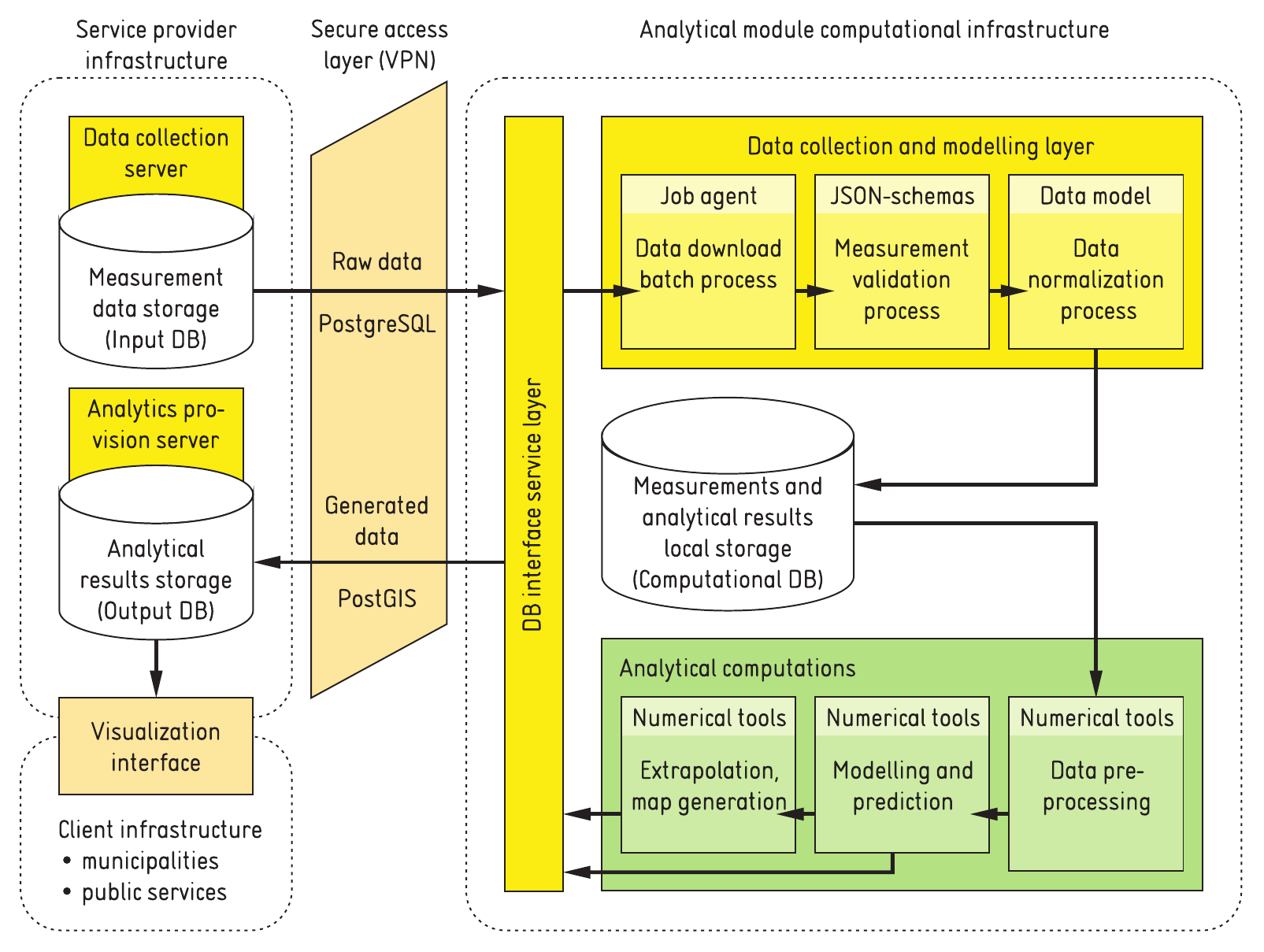}
\caption{Computational and database concept of the analytical module.}
\label{fig:architectural-conception}
\vspace{-2mm}
\end{figure}

The analytical module is a fully cloud-based {\em Software as a Service} (SaaS) solution, which runs on several virtual nodes in a cloud 
infrastructure. 
Figure \ref{fig:architectural-conception} illustrates the computational and database architecture of the module. The {\em scalability} and {\em configurability} of the platform are guaranteed thanks to the availability of numerous parameters for analytics generation and customization. Typically, parameters such as map origin, length and bearing, prediction confidence bounds, shift and horizon can be changed.
Newly added virtual nodes can be on-demand allocated and clustered, in compliance with new computational requirements eventually emerging.

The analytics generation process was essentially implemented as a parallel computing constellation in the cloud, by exploiting clusters as replicated resource templates (levering the powerful concept of cloud-node commodity), which enabled a perceivable performance-to-resource processing mechanism for generating the analytics, and provided also high flexibility and technology tracking. New sensory data are constantly downloaded from the input database and then processed, normalized and stored into a different, analytics suitable output database. Data are validated applying a cascading approach, and normalized against formally specified representation models (JSON 
and PostGIS).
Data relative to new sensor-boxes and measurement dimensions could be dynamically recognized and modelled by the analytical module, by adapting and augmenting the underlying database model. The analytics calculated for the specified quantities at discrete points of time and discrete spatial points corresponded to specific geographical locations. Generated analytical data are finally replicated (in PostGIS compliant format) both on the virtual nodes of the module and on the output
database.

\subsection{Geographic Information System Based Analytics}
\label{subsec:gis}
Short-term forecasts and smoothed maps are generated producing PostGIS-compliant spatial information. PostGIS is a spatial database extension for PostgreSQL
object relational DBMS, with support for geographic objects, allowing location queries and providing functions, operators, and index enhancements inherent to these spatial types. 
The following are the PostGIS-supported geometric objects (data-types) primarily leveraged in the analytical module: {\sc Point(X,Y,Z)} for three-dimensional GPS positions; 
{\sc PolyhedralSurface}
and 
{\sc Point(X,Y,Z)} for smoothed intensity maps; 
{\sc LineString} 
for forecasts and confidence (prediction) regions, and 
{\sc MultiLineString}
for upper and lower confidence boundary ranges.

\subsection{System Persistence and Storage Requirements}
The
persistence service layer has been designed with interoperability and robustness in mind, by orchestrating context-specific modular services (measurement data download, relational transformation and normalization of data sequences), and by computing and storing GIS-compliant information for forecasts and intensity maps. Database testing included structural and functional tests: the first focused on elements validation of the repository data (primarily used for storage and that are not allowed to be directly accessed and manipulated by the end users, i.e. schemas, tables, stored procedures and so forth), whereas functional testing encompassed activities aiming at proving the transactional and operational soundness of the analytical process and its consistency against the application requirements. 
An initial rough estimation of measurement data showed that one sensor box (of the initial 250 preliminarily designed) might approximately provide 80 MB\,/\,month of raw sensor data (as measured on the wire). This means 20 GB\,/\,month, 240 GB\,/\,year, or about 7 MB\,/\,15 min. (database storage overhead, indexes, etc. not considered here). 

The real growth trend in measurement data evidenced a linear tendency, with a bounded weekly increase in the database size of approx. 180 MB for data download (9.2 GB for a whole year, if tendency remains unaltered for all of the dimensions currently measured). The growth trend for the output database evidenced on the other hand an increase of approx. 2030 kB in maps storing and 1580 kB for forecasts one, always related to one week of generated analytical outputs. Concluding, this means that approx. 103 MB for maps and 80 MB for forecasts are required for storing all of the generated results throughout a whole year of analytical computations (always assuming no change in the application logics and 
data representations).

%%%%%%%%%%%%%%%%%%%%%%%%%%%%%%%%%%%%%%%%%%%%%%%%%%%%%%%%%%%%%%%%%%%%%%%%%%%%%%%%%%%%%%%%%%%%%%%%%%%%%%%%%%%%%%%%%%%%%%%%%%%%%%%%%%%

\section{Pre- and Post-Processing}
After reading out new measurements from the input database, the module first pre-processes the data. Pre-processing is a crucial step, as the raw observations usually suffer from various problems which prevent the immediate application of statistical techniques \cite{Ljung1999}. 
Such problems with the measurements 
could be: high-frequency disturbances (above the interesting frequencies); low-frequency (possibly periodic) disturbances; outliers and bursts; missing measurements; inconsistent data; asynchronous observations; drifts and offsets. 

The role of post-processing is mainly to transform back the outputs of the analytical module to the original coordinate system and to fit geometrical objects to the results.

\subsection{Pre-Processing}
Now, we briefly summarize how the analytical module transforms the raw measurements into a cleaned dataset.
\smallskip

\begin{figure}[t]
	\begin{center}
		\includegraphics[width=\columnwidth]{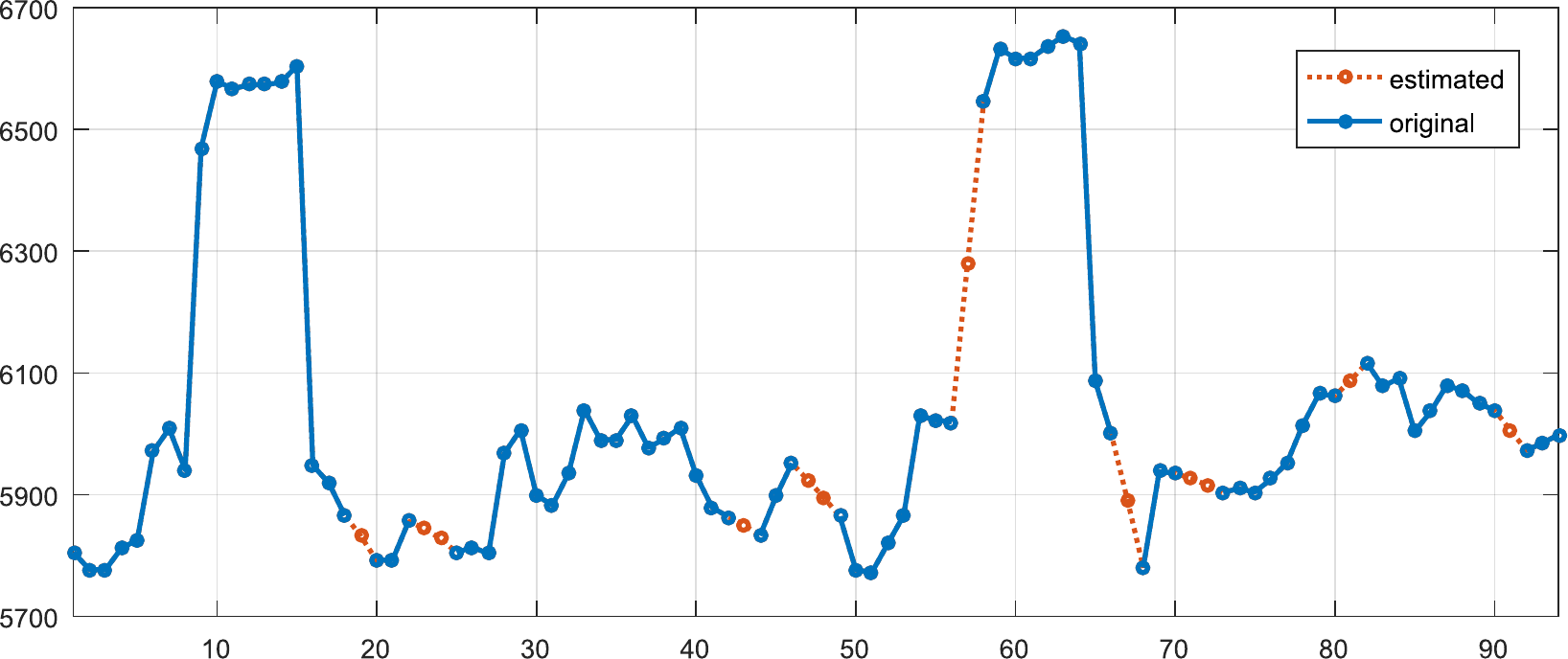}
		\caption{Estimating missing particulate matter measurements. The blue dots show the available data, while the red points are the estimates.}
		\label{fig:missing}
		\vspace{-3mm}
	\end{center}
\end{figure}

\begin{enumerate}
	\item{\em Filtering Outliers.} Outliers are seriously corrupted data (typically resulting from physical errors) which can considerably mislead and bias the applied statistical methods. As their absolute values are often much larger than the values coming from ``normal'' working conditions, it is not advised the leave their removal to the ``smoothing'' process, as they can drastically change the smoothed value. A problem with outliers is that they typically have some delayed effect, as well, e.g., the system may only slowly return to its normal working condition, thus simple thresholding (e.g., the process itself or its derivative) is not enough to eliminate them. The analytical module applies a Hampel filter \cite{liu2004line}, 
to remove outliers.
It is based on a sliding window and tests how much the center of the window differs from the median of the window. If it differs by more than some constant times the standard deviation, then the center is classified as an outlier.
\smallskip
 
	\item{\em Discretization.} In order to apply the machinery of time-series analysis, which assumes discrete-time processes which were obtained with a constant sampling rate \cite{Ljung1999}, the data are discretized. A simple approach to do so is to set a large enough (virtual) sampling rate 
	and identify the value of the signal with the average measurements in each corresponding interval. Naturally, this step may result in a time-series with missing measurements. \smallskip
	\item{\em Standardization.} The module then centers and normalizes the data.
	The scaling is done mainly for numerical stability, while centered data are often presupposed by various statistical methods. Therefore, after discretization, the data is transformed to ensure that its (empirical) mean is zero, while its (empirical) standard deviation is one. \smallskip
		\item{\em Missing Information.} The problem of missing data points (w.r.t.\ the discretized time axis) is handled by estimating them with an initial (crude) time-series model \cite{Ljung1999}. This initial model is typically either a simple (low order) auto-regressive (AR) or a moving average (MA) model. First,
the model is identified based on the available data, then the missing values are estimated using the model. This process may also be repeated iteratively, in order to improve the solution \cite{Ljung1999}. 
Figure \ref{fig:missing} illustrates the results for the case of missing particulate matter measurements.\smallskip

\item{\em Smoothing.} The measurements are always corrupted by noise whose effect should be reduced to improve the solution. The analyitcal module smooths the data by removing the high- and low-frequency disturbances via computing the (circular) convolution of the signal with a suitably scaled $sinc$ function. This is, of course, equivalent to first transforming the signal to the frequency domain (with the Fourier transform), multiplying it with a rectangle function, then returning to the time domain \cite{korner1989fourier}.\smallskip

\item{\em Typical Values.} Finally, the periodic average behavior of the processes are computed. It is common that stressors (such as noise, temperature, and UV irradiation) have a quasi-periodic nature, e.g., their daily progresses have some recurring patterns. Having an estimate of these patterns is very useful for forecasting. Therefore, we compute the average values
for each timestep of the day. For example, if the stepsize is one hour, we compute a typical value for each hour of the day. A sliding window (e.g., one month) is used, hence, only the most recent values are considered.
The sliding window guarantees that seasonal changes are automatically taken into account, thus, the typical values smoothly change over the year.
\end{enumerate}

\subsection{Post-Processing}
Post-processing is only initiated after the forecasts, maps, and reliability estimates were computed. The computed quantities are first transformed back to the original coordinate system of the measurements by inverting the scaling and centering done during pre-processing. Then, geometrical objects are fitted to the data, particularly {\sc LineString}-s and {\sc PolyhedralSurface}-s, which are then recorded in the output database using a GIS representation. This approach allows flexible queries,
for example, based on the GIS representation the database system can answer queries asking for the integral of a 
quantity over a given (time or space) domain.

%%%%%%%%%%%%%%%%%%%%%%%%%%%%%%%%%%%%%%%%%%%%%%%%%%%%%%%%%%%%%%%%%%%%%%%%%%%%%%%%%%%%%%%%%%%%%%%%%%%%%%%%%%%%%%%%%%%%%%%%%%%%%%%%%%%

\section{Forecasting}

In this section we turn our attention to the problem of estimating time-series models, applying them to generate short-term forecasts with corresponding prediction regions. We will assume that we have a pre-processed dataset, $\{x_t\}$, i.e., a finite sequence of cleaned observations (e.g., without any outliers or missing data) as well as a sequence of typical values, $\{u_t\}$, that we use circularly, i.e., we treat them as a periodic sequence and hence, for any integer $t$, $u_t$ is well-defined.

\subsection{Estimating Time-Series Models}
The problem of estimating dynamical models from experimental data is also called {\em system identification} and has a rich literature 
\cite{Ljung1999}. 
The analytical module applies {\em parametric estimation} methods that is it assumes that our model class is parametrized by a finite dimensional vector, $\theta \in \mathbb{R}^d$. Thus, finding a suitable model is equivalent to finding a parameter corresponding to the model which best fit the data.

During the project a number of time-series models were tested, including Box-Jenkins, Hammerstein-Wiener, kernel regression, multilayer perceptron, and wavelet based approaches. However, it was found that the targeted stressors can be very well represented by standard {\em auto-regressive exogenous} (ARX) models, in case the exogenous components are chosen well.

ARX models are defined as follows 
\cite{Ljung1999}
$$A(z; \theta)\,x_t \, \triangleq\, B(z; \theta) \, u_t + n_t,$$
where $x_t$ is the output, $u_t$ is the input and $n_t$ is the 
noise at time $t$; and $A(z; \theta)$ and $B(z; \theta)$ are polynomials in the backward shift operator, $z^{-1}$, i.e., $z^{-i}x_t \triangleq x_{t-i}$, that is
\begin{eqnarray*}
A(z; \theta) &\,\triangleq\,& 1-a_1z^{-1}-a_2z^{-2}-\dots-a_pz^{-p},\\
B(z; \theta) &\,\triangleq\,& b_0z^{0}+b_1z^{-1}+\dots+b_{q-1}z^{-q+1},
\end{eqnarray*}
where parameter vector $\theta \in \mathbb{R}^{p+q}$ contains the constants $\{a_i\}$ and $\{b_j\}$. We have also found that using an ARX structure with orders $p=2$ and $q = 2$ work well for all targeted stressors, as demonstrated by Table \ref{tab:rmse} in Section 
\ref{valid-models}. The exogenous inputs $\{u_t\}$ were the typical values calculated from the last $30$ days of measurements of the specific sensor group.

Using ARX models is numerically cheap, they only require simple matrix arithmetics, and thus can also have a direct hardware implementation. Even {\em fitting} ARX models
to the pre-processed data requires basically a matrix inversion, as it is based on the standard {\em least-squares} approach, which has an analytical solution \cite{Ljung1999}. 
Thus, ARX based models scale well.

\subsection{Bootstrapping: Estimating and Generating Noises}
Having a time-series model at hand, we proceed with simulating the future behavior of the system, in order to compute short-term forecasts and prediction regions for the stressors. However, we also need a model of the noise driving the system to be able to simulate the process. The prediction errors, $\{\varepsilon_t(\hat{\theta})\}$, of the least-squares estimate, $\hat{\theta}$, defined as
$$\varepsilon_t(\hat{\theta}) \, \triangleq\, A(z; \hat{\theta})\,x_t - B(z; \hat{\theta}) \, u_t,$$
can be seen as estimates of the noise driving the process.

Instead of assuming that the noises have specific known distributions (e.g., Gaussian), we directly use the empirical distribution function of the prediction errors to generate new noise instances, which approach is often referred to as {\em bootstrap} \cite{efron1994introduction}. The {\em empirical distribution function} (EDF) is 
\cite{schervish2012theory}
$$\widehat{F}(x; \hat{\theta}) \,\triangleq\, \frac{1}{n} \sum_{i=1}^n \mathbb{I}(\,\varepsilon_t(\hat{\theta}) \leq x\,),$$
where $\mathbb{I}(\cdot)$ is an indicator function, i.e., its value is $1$ if its argument is true and $0$ otherwise. It is known (cf.\ the Glivenko-Cantelli theorem) that as the sample size increases the empirical distribution function will {\em uniformly converge} to the true cumulative distribution function, assuming the data is independent and identically distributed (i.i.d.) \cite{schervish2012theory}.

It is important to note that we calculate a separate EDF for each interval of the day. For example, if the step-size is $60$ minutes, we have $24$ EDFs. We do so, because very often the fluctuations of the stressor processes depend on time.

\begin{figure}[t]
\begin{center}
\includegraphics[width=\columnwidth]{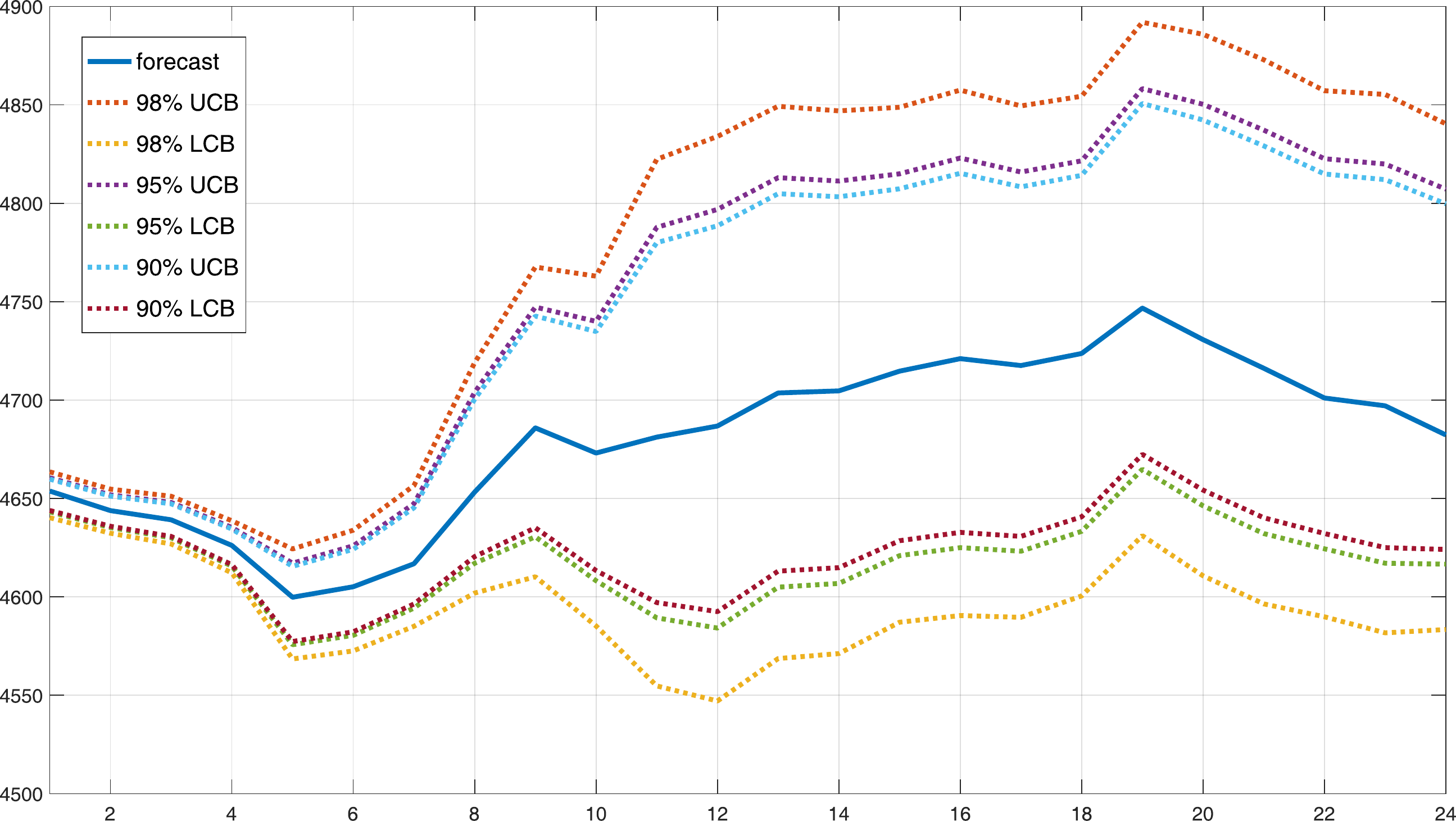}
\caption{Forecast and prediction regions for particulate matter levels.}
\label{ref:forecast}
\end{center}
\vspace{-3mm}
\end{figure}

\subsection{Monte Carlo Forecasts and Prediction Regions}
After the distribution of the noise was estimated, Monte Carlo simulations \cite{gilks2005markov} can be carried out, using the last values of our observations as initial states and randomly generated noise according to the identified noise distributions (the EDFs for the specific times of the day), to generate simulated trajectories. Then, approximate upper [lower] prediction bounds can be calculated from the simulated trajectories by finding the smallest [largest] sequence that is larger [smaller] than a given confidence percentage, e.g., $95$\,\%, of the trajectories. Short-term forecast can also be computed from the Monte Carlo simulations by calculating the mean of the trajectories.

Note that directly applying the linearity of the ARX models to get forecasts may results in forecasts incompatible with the prediction regions, thus, computing the forecasts from the Monte Carlo trajectories is applied by the module. 

A 24-hour forecast of particulate matter levels (with 1-hour stepsize) accompanied by predictions regions with various lower and upper confidence probabilities (90\,\%, 95\,\%, 98\,\%) is illustrated by Figure \ref{ref:forecast}. The prediction regions can help to evaluate the reliability of the obtained short-term forecast.

%%%%%%%%%%%%%%%%%%%%%%%%%%%%%%%%%%%%%%%%%%%%%%%%%%%%%%%%%%%%%%%%%%%%%%%%%%%%%%%%%%%%%%%%%%%%%%%%%%%%%%%%%%%%%%%%%%%%%%%%%%%%%%%%%%%

\section{Smoothed Maps}
\label{sec-maps}
A common requirement both for aggregated measured data and predictions is their visualisation in the geographical context, typically via map surfaces interpolated\,/\,extrapolated based on point-wise values. While such functionalities can be migrated to visualisation interfaces in a commercial roll-out of the system (\eg to cut down excess data storage needs), the experimental version of the analytical module does include a number of map generation options for test purposes.

\subsection{Map Structure}
The smoothed maps generated by the analytical module define raster points of a rectangle projected over a geographical area, with the location of a corner point and the bearing (rotation angle) of the map being specified, in addition to the edge lengths and the number of raster units in each direction.
Hence, the maps can have an anisotropic resolution if needed.
Each raster point of a generated map is specified by (1)~its geographical coordinates, (2)~value of the interpolated\,/\,extrapolated quantity (referred to as \emph{value point}), and (3)~an estimated measure of reliability of the given value (referred to as \emph{reliability point}). 
Recall that in the post-processing phase, a {\sc PolyhedralSurface} is fitted to the value points (and another surface to the reliability points) and stored using GIS representation in the output database.

\subsection{Injecting Auxiliary Data}
\label{sec-virt-sens}
In some cases, injection of auxiliary data is advantageous to create more plausible maps. {\em Virtual sensor} values at additional locations can be derived from raw readings relying on field knowledge (\eg various stressors being consistently channelled or blocked by constrained spaces), or fixed values based on preliminary assessment (\eg areas largely shielded from external influence
by buildings). Depending on the type of map generation, virtual sensors can have local influence with finer control of local modifications (at the cost of more virtual sensors being required to tune extended areas), or can exert influence on the entire area within their convex hull, if for example a neighbourhood-based method is used.

\subsection{Interpolation and Extrapolation}
The WSN in question deploys stationary sensors only, each having fixed geographical coordinates---these are assumed to remain constant, just as unique identifiers and further fixed parameters of the sensors. The location of the sensors does, however, not correspond to raster points of a rectangular grid---therefore, measurements and predictions assigned to sensor locations are considered \emph{scattered data}, from which the values for the map raster points are obtained by \emph{scattered data interpolation or extrapolation} which calculates a scalar value $z$ for a query point $Q$ based on the $\{z_i\}$ values of the scattered points $\{P_i\}$. In this case, $\{P_i\}$ are the sensor locations, while each map raster point acts as a query point $Q$. 

Several methods are known to work efficiently with data of geographical relevance (for which, thus, certain consistency characteristics can be assumed), and some of them are capable of extrapolating both through time and space. The experiments reported in the paper followed a two-step approach instead:
\medskip
\begin{enumerate}
\item{}
\begin{enumerate}
\item{Quantities measured or predicted over a specified time interval were sorted by the unique identifier of the corresponding sensor.}
\item{If no measured value was encountered for a given sensor, the search could optionally extend to neighbouring intervals in an attempt to find valid data.}
\item{If several measurements or predictions were found for a given sensor, the median (optionally, a windowed average around the median) is calculated and assigned to the sensor as if it were a single measured or predicted value, so that all sensor locations $P_i$ have only one scalar vale assigned: $P_i \mapsto z_i~\forall i$.} 
\end{enumerate}\medskip
\item{}
\begin{enumerate}
\item{If the selected method can only interpolate and the convex hull of the scattered points does not cover the entire map, surrogate values (corresponding to {\em virtual sensors}) are calculated for the map corner points by an extrapolation method of choice. In subsequent calculations, these additional points are treated in the same way as values assigned to actual sensor locations.}
\item{The values for the map raster points are calculated by spatial-only scattered interpolation\,/\,extrapolation.}
\end{enumerate}
\end{enumerate}
\medskip
Several interpolation\,/\,extrapolation methods were tested, primarily with measured data as these are by nature more challenging for the robustness and fault tolerance of the map calculation methods. The list below gives a brief summary of the algorithms tested, as well as our experience with the data.

\emph{Nearest neighbour interpolation} assigns the $z$ value of the data point closest to $Q$ to the query point $Q$: 
\begin{eqnarray*}
z(Q)\,\,\triangleq\,\,z_{\arg \min d(Q,P_i)},
\end{eqnarray*}
where $d(Q,P_i)$ is typically the Euclidean distance of the query point and the data point in question. While the resulting rough terrain has distinct plateaus that prove inferior in a number of applications, an advantage of the nearest-neighbour method is that it does not restrict itself to interpolation in the convex hull of scattered points but is capable of extrapolating over an entire Euclidean space. Also, the method is not prone to producing overshoot---the latter may cause undesired peaks, \eg for sensors that are located close to each other but deliver greatly different values. Also, the effect of remote sensors is essentially blocked by the first ring of scattered points around the query point. This makes all neighbourhood-based interpolations particularly suitable for deployment where terrain properties or artifacts (walls, massive vegetation, etc.) exhibit the same blocking nature in reality.

\emph{Natural neighbour interpolation} returns a \emph{weighted average} of $\{z_i\}$ values based on area occupation ratios in Voronoi regions. Let us assume the same set of scattered data points $\{P_i\}$ with $P_i\mapsto z_i~~\forall i$, and let us consider the Voronoi cells around each scattered data point. Insert the query point $Q$, a new Voronoi cell is formed around it, occupying a certain area of each neighbouring cell, and the weight $w_i$ calculated for each neighbouring data point $P_i$ is some function of this occupied area. The are different methods to obtain $\{w_i\}$, see the broad overview of \cite{bobach2006comparison} for examples.

Once the weights, $\{w_i\}$, are obtained, $z(Q)$ is formed as the weighted average of $\{z_i\}$ values.
Natural neighbour interpolations yield a 
better $z$-terrain than nearest-neighbourhood methods, and are likewise free of overshoot and block the effect of remote scattered points. Most natural neighbour interpolations, however, are restricted to the convex hull of $\{P_i\}$, and therefore may require extra map corner points (virtual sensors) with extrapolated $z$-values.

\begin{figure}[t]
\begin{center}
\includegraphics[width=\columnwidth]{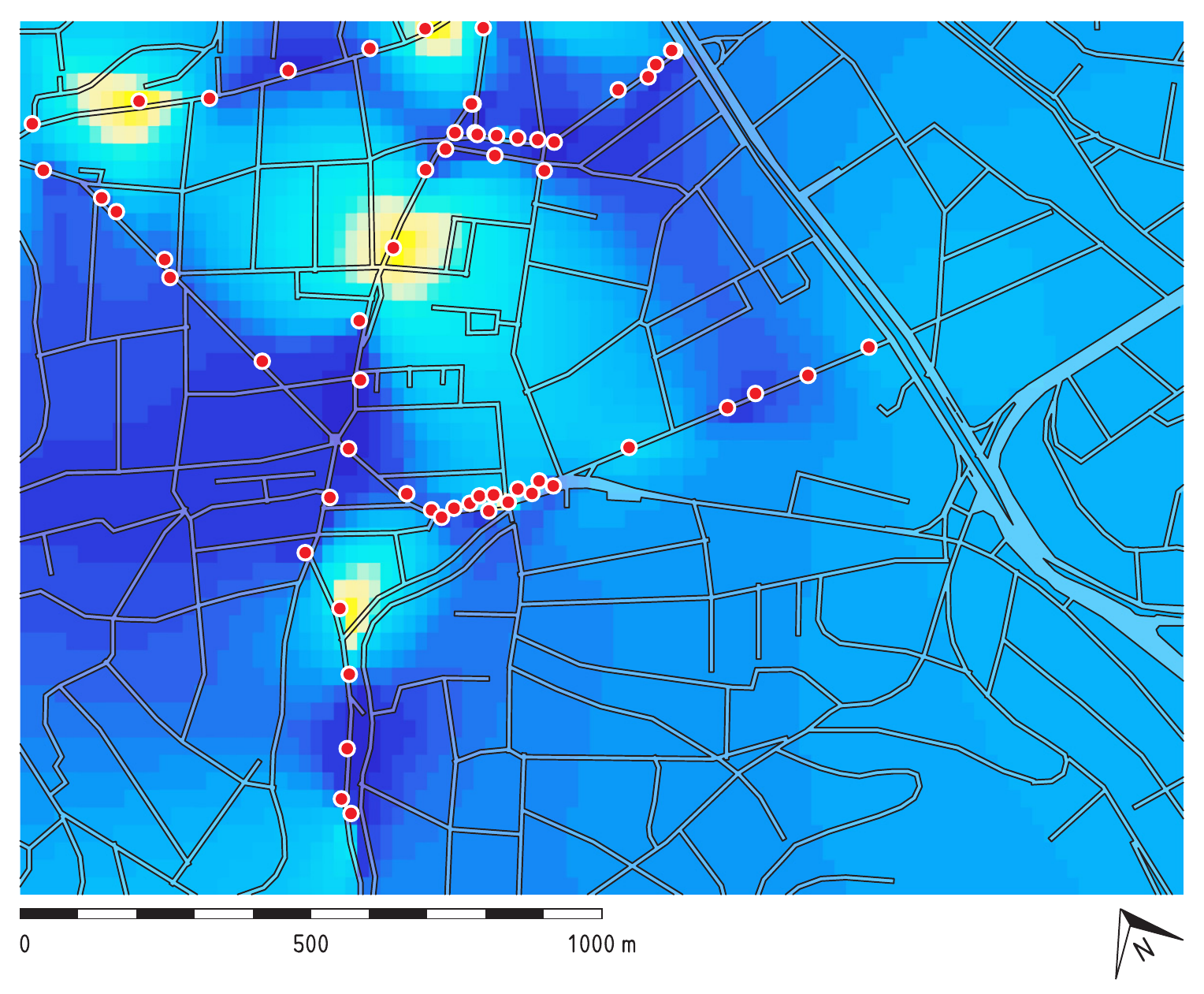}
\caption{Example of a hourly generated particulate matter map constructed using {\em natural neighbour} interpolation. For full map coverage, map {\em corners} were extrapolated using exponential distance metrics with 
$1/d^2$
and inserted into the sensor data set, as new (virtual) sensors. 
The red dots denote the locations of the original sensors. Note the high values rising far above average at some locations---this is unavoidable, even with properly calibrated sensors, due to the nature of certain urban activities, \eg construction and demolition sites, or large vehicles idling for an extended period of time.}
\label{fig:mapx_natural}
\end{center}
\vspace{-3mm}
\end{figure}

\emph{Linear scattered data interpolation} is, in a geometrical sense, equal to fitting planes onto the terrain sampled by the scattered points. In a technically simple case, the argument space (\ie the space in which the $\{P_i\}$ points are located) is triangulated with the help of segments connecting neighbouring scattered points. These are often referred to as  \emph{Delaunay edges}
\cite{preparata2012computational}, 
and are dual to the Voronoi cell boundaries. Linear interpolation within these triangles is then a weighted average of the $\{z_i\}$ values assigned to the three $\{P_i\}$ points at the vertices of the triangle. This ensures that (1)~neighbouring triangles of the $z$-terrain fit to each other forming a continuous (and piecewise linear) function, and (2)~no overshoot is experienced as the extrema of $z$ are at scattered data point locations that are vertices of the triangles. Methods fitting planes over a larger number of scattered points (\eg for surface reconstruction in geometrical reverse engineering) are not necessarily free of overshoot
\cite{chivate1995review}. 
Similarly to neighbourhood-based interpolation, the effect of remote scattered points is blocked. Linear interpolation is, however, restricted to the convex hull of $\{P_i\}$. 

\emph{Interpolation with distance metrics} is a computationally efficient method of scattered data interpolation.
Here, the $z$ value for the query point $Q$ is interpolated by a weighted average, where weights $\{w_i\}$ are based on some distance metric function that monotonously decreases with the---mostly Euclidean---distance $d$ between $Q$ and $\{P_i\}$, that is
$$
z(Q)\,\,\triangleq\,\,\frac{\sum_i w_iz_i}{\sum_i w_i},\hspace{3mm} \textrm{where} \hspace{3mm}w_i\,\,\triangleq\,\,f\left(d(Q,P_i)\right).
$$
One of the most commonly used distance metric functions is
$f(d)\,\triangleq\,\frac{1}{d^n}$, where $n$ is typically $2$, or some other relatively low exponent. Distances close to $d=0$ are detected and truncated by an exception handling mechanism. With increasing $n$, the resulting interpolated $z$-terrain is converging to the nearest-neighbour plateaus; in fact, nearest-neighbour interpolation can be regarded as the special case of $n=\infty$. 

\begin{figure}[t]
\begin{center}
\includegraphics[width=\columnwidth]{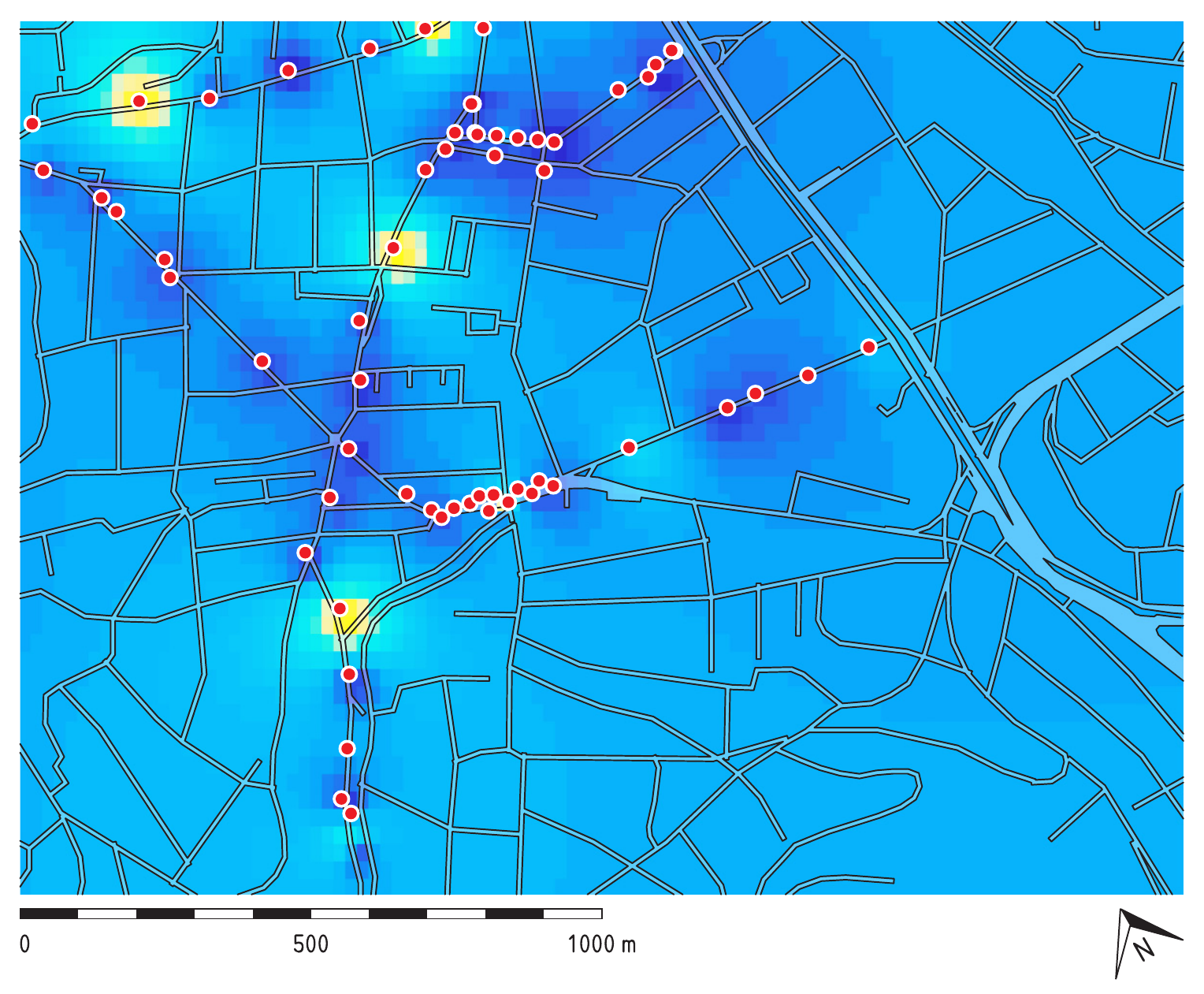}
\caption{Example of a map extrapolated using {\em exponential distance} metrics with $f(d)=1/d^2$. The scattered data are identical to those used in Fig.\ \ref{fig:mapx_natural}}
\label{fig:mapx_exp20}
\end{center}
\vspace{-3mm}
\end{figure}

While distance metric-based interpolation can be perturbed by vast outliers of sensors further away, it still does not produce overshoot, and can deliver values outside the convex hull of scattered data points. In the current implementation of the analytical module, distance metric-based interpolation with $n=2$ is used (see also Figure \ref{fig:mapx_exp20}).

\emph{Spline and curve fitting interpolation approaches} fit a parametric, usually polynomial-based, function over scattered data points while optimizing various criteria on approximation error, standard deviation, performance measures related to derivatives, or physically funded measures as friction of a bent wire\,/\,sheet passing through specific corner points. Splines and related interpolation approaches form a populous class, and are subject to intense research to date
\cite{mitas1999spatial,lee1997scattered}. 

A major advantage of these approaches is the wide spectrum of possibilities of tailoring the interpolation 
to the needs of the given application. For example, smooth appearance can be achieved at the cost of giving up the strict adherence to (noisy) scattered data, or interpolation artefacts (\eg step-like changes in temporal processes calculated over roughly-spaced time windows) can be reduced 
\cite{appice2013using}. 
Some methods, \eg \emph{kriging} \cite{stein2012interpolation}, assume the knowledge of an underlying model governing changes of the $z$-terrain. 
Others use, \eg \emph{radial basis functions} (RBF) and assemble the interpolated terrain from radially symmetrical (typically Gaussian) functions 
\cite{powell1987radial}.

Experiments with spline interpolation shown 
overshoots in the presence of noisy 
data, which could be explained by the sensitivity of such methods to inaccurate derivative estimates.

\subsection{Reliability Estimates}
Reliability points express a measure of certainty for the values estimated at the map raster points. In a graphical map, reliability measure can be visualised, \eg as colour intensity, or be shown with icons corresponding to quantised measures. Backed by a longer period of operation and a large corpus of data, reliability measures can be a function of several features extracted from sensor data or location:
\begin{itemize}
\item{Distance of the query point from the scattered data point;}
\item{Statistical properties of readings from the given sensor;}
\item{Consistency with data yielded by nearby sensors;}
\item{Known influence of artefacts and other terrain properties.}
\end{itemize}
The current implementation uses distance-based reliability,
\begin{eqnarray*}
r(Q)\!\!&\triangleq&\!\!\max_i g_{\textrm{xtype}}\left(d(Q,P_i)\right),\\
g_{\textrm{xtype}}(d)\!\!&\triangleq&\!\!e^{-\frac{d^2}{2c_{\textrm{xtype}}}},
\end{eqnarray*}
where $c_{\textrm{xtype}}$ is an empirically obtained constant for each type of physical quantity (stressor) measured.

%%%%%%%%%%%%%%%%%%%%%%%%%%%%%%%%%%%%%%%%%%%%%%%%%%%%%%%%%%%%%%%%%%%%%%%%%%%%%%%%%%%%%%%%%%%%%%%%%%%%%%%%%%%%%%%%%%%%%%%%%%%%%%%%%%%

\smallskip
\section{Validation}
\begin{figure}[t]
	\begin{center}
		\includegraphics[width=\columnwidth]{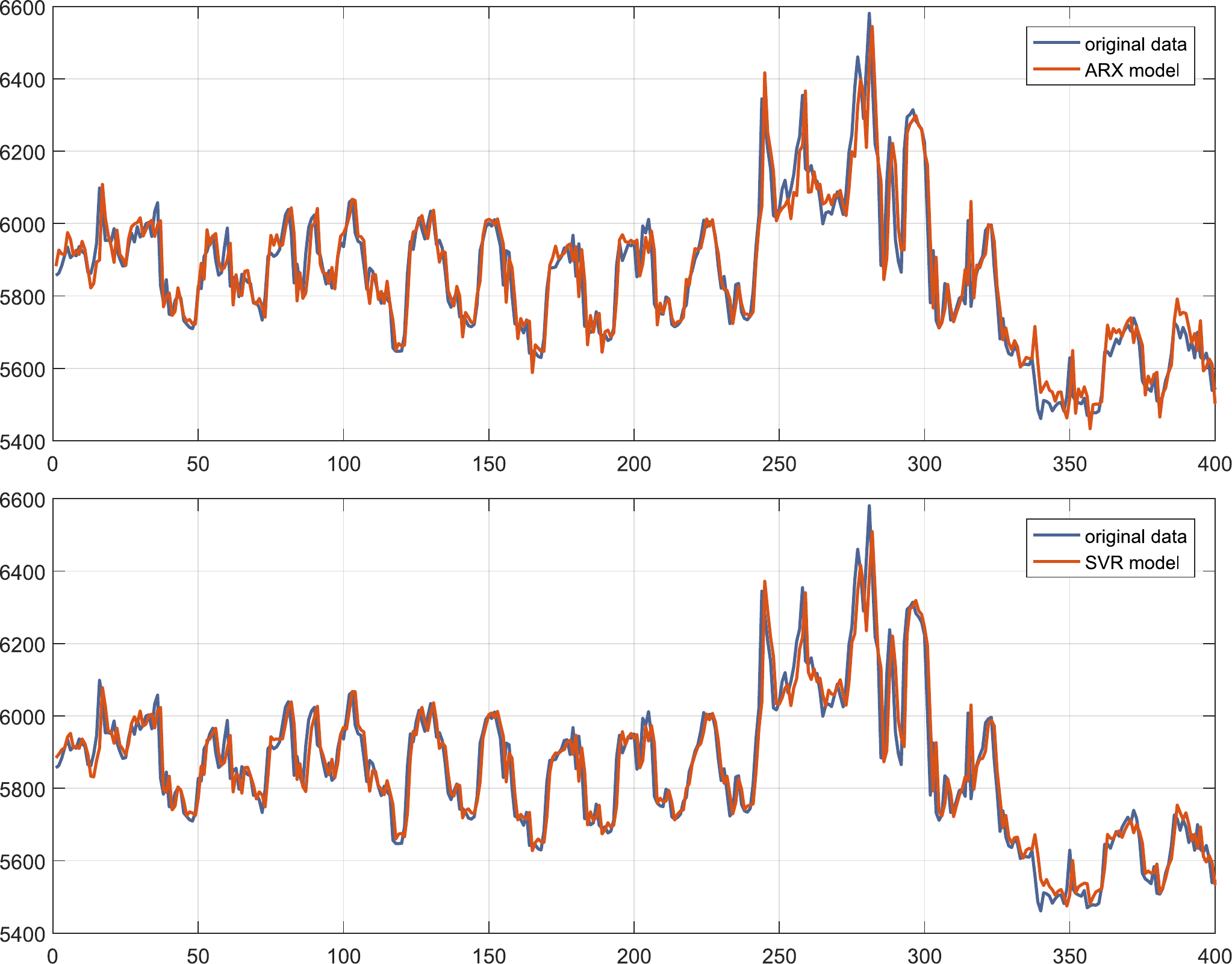}
		\caption{Modelling particulate matter with ARX and SVR models. The blue lines show the original data, while the red ones are based on the models.}
		\label{fig:dust}
	\end{center}
	\vspace{-3mm}
\end{figure}

In this section we present results about the validation of the forecasting and map generation capabilities of the module. 

\subsection{Validation of the Time-Series Models}
\label{valid-models}
The efficiency of  the suggested ARX models was verified by testing their performances on a separated validation (test) dataset and by comparing their results with 
that of popular (nonparametric, nonlinear) {\em support vector regression} (SVR) techniques \cite{scholkopf2001learning}. ARX and SVR models for various stressors were computed  for the period between 1st May, 2016 and 31st May, 2016. The available data were split into a training (estimation) dataset and a test (validation) dataset, in a way that we used 2\,/\,3 of the data as training data and 1\,/\,3 as test data. The data were pre-processed, 
which should be taken into account when interpreting the root-mean-square-errors.

As it can be observed from Table \ref{tab:rmse}, both model types fit the data well and have good generalization properties. Moreover, the standard (linear, parametric) ARX models achieved very similar fit and prediction results on these data to the more complicated SVR models. This phenomenon can be explained, for example, by the extensive pre-processing and the relatively slow dynamics of the processes at hand. This indicates that, {\em for these specific kinds of data}, ARX models should be used for the analytical module, as they have several advantages. For example, they are simpler to represent (require less memory), easier to fit and calculate with, easier to interpret, and moreover, their Vapnik-Chervonenkis (VC) dimension \cite{vapnik1998statistical} is lower, thus they have better (theoretical) generalization properties. 

Fitting ARX and SVR models is also illustrated by Figure \ref{fig:dust}. It shows particulate matter data, whose modelling was more difficult than modelling other stressors (cf.~Table \ref{tab:rmse}).

{\renewcommand{\arraystretch}{1.2}
	\begin{center}
		\begin{table}[b]
			\vspace*{-3mm}
			\caption{Root-mean-square-errors of ARX and SVR predictions}
			\centering
			\begin{tabular}{|l||c|c||c|c|}\hline
				\multicolumn{1}{|c||}{}&\multicolumn{2}{c||}{ARX Models}&\multicolumn{2}{c|}{SVR Models}\\ \hline
				Stressor &\!\! Estimation\!\! &\!\! Validation\!\! &\!\! Estimation\!\! &\!\! Validation\!\! \\ \hline\hline
				particulate matter        &  0.1299 & 0.1816 & 0.1312 & 0.1763 \\ \hline
				temperature               &  0.0688 & 0.0828 & 0.0677 & 0.0852 \\ \hline
				UVB irradiation            &  0.0880 & 0.1050 & 0.0861 & 0.1043 \\ \hline
				ambient light             &  0.1073 & 0.1249 & 0.1055 & 0.1236 \\ \hline	
				air pressure              &  0.0273 & 0.0346 & 0.0280 & 0.0349 \\ \hline	
				relative humidity         &  0.0967 & 0.1093 & 0.0965 & 0.1126 \\ \hline	
				carbon monoxide           &  0.0553 & 0.0747 & 0.0558 & 0.0908 \\ \hline
			\end{tabular}
			\label{tab:rmse}
		\end{table}
\end{center}}

\vspace{-8mm}
\subsection{Validation of the Smoothed Maps}
Now, we turn our attention to the validation of the smoothed map generation process.
The test period was from 1st May, 2016 till 31st July, 2016. Maps were generated for each hour of the test period; the observations were pre-processed, i.e., (1) {\em outliers} were removed, (2) the measurements were {\em normalized}, and (3) their {\em medians} (w.r.t.\ the given hour) were computed.

The proposed map generation methods were tested (for each air quality related stressor and each hour of the test period) by {\em leave-one-out cross-validation}, i.e., for each sensor its median measurement was compared with the estimate coming from the map generated by the medians of all other sensors. This process was repeated for all sensors (and all hours) and the corresponding {\em root-mean-square-errors} were computed. The results for the two best methods, i.e., natural neighbour and inverse-square distance based, are presented in Table \ref{tab:map-rmse}.

It can be observed that for some stressors, like temperature and relative humidity, the measurements were very well extrapolated, while for others, such as ambient light and particulate matter, the errors were higher, but still acceptable. This latter phenomenon can be explained by the fact that two spatially close sensors may have very different light or dust conditions. This also points in the direction that the topological relationships between the sensors should be identified, in order to improve the constructions (then, e.g., the maps could be refined by virtual sensors discussed in Section \ref{sec-virt-sens}). This is a possible future direction. Nevertheless, Table \ref{tab:map-rmse} confirms that the maps currently generated extrapolate efficiently.

{\renewcommand{\arraystretch}{1.2}
	\begin{center}
		\begin{table}[t]
			\caption{Root-mean-square-errors of map generation methods}
			\centering
			\begin{tabular}{|l||c||c|}\hline
				Stressor & \hspace{3mm} Natural neighbour \hspace{2mm} & Inverse-square distance \\ \hline\hline
				particulate matter       & 0.094427  &  0.086928 \\ \hline
				temperature                & 0.023348 & 0.023532  \\ \hline
				UVB irradiation           & 0.100946 & 0.099960 \\ \hline
				ambient light              & 0.097892 & 0.096710  \\ \hline
				air pressure               & 0.052853 & 0.050374  \\ \hline	
				relative humidity       & 0.026321 & 0.026674 \\ \hline	
				carbon monoxide      & 0.205682  & 0.173037 \\ \hline
			\end{tabular}
			\label{tab:map-rmse}
		\end{table}
\end{center}}

%%%%%%%%%%%%%%%%%%%%%%%%%%%%%%%%%%%%%%%%%%%%%%%%%%%%%%%%%%%%%%%%%%%%%%%%%%%%%%%%%%%%%%%%%%%%%%%%%%%%%%%%%%%%%%%%%%%%%%%%%%%%%%%%%%%

\vspace{-8mm}
\section{Conclusions}
The paper presented a smart city prototype experiment in which smart sensor boxes (each containing a group of different sensors, a battery and a transceiver) were installed on the public lighting system in District XII of Budapest, Hungary. 

The sensors primarily measure air quality and traffic related quantities, and send their measurements through a public mobile communication network to a dedicated database for raw sensor data. The data-processing is done by a separate cloud-based analytical module that periodically generates short-term forecast and smoothed maps, both accompanied by reliability estimates. The results of the module are stored in an output database, using geometric representations allowing flexible queries, which can constitute a basis for public services.

It was shown that, after pre-processing, air quality related stressors can be well modelled with ARX type models, forecasts can be created by bootstrap and Monte Carlo  techniques, while for map generation natural neighbour interpolation and inverse-square distance metrics provided the best results.

The potential joint analysis of stressors could help to study their interdependencies, or to make better forecasts and maps, or even to improve the estimation of missing measurements. 

Possible future research directions are to study methods that can evaluate multiple stressors simultaneously, provide traffic specific analyses and refine maps based on topological data.\vspace*{5mm}

%%%%%%%%%%%%%%%%%%%%%%%%%%%%%%%%%%%%%%%%%%%%%%%%%%%%%%%%%%%%%%%%%%%%%%%%%%%%%%%%%%%%%%%%%%%%%%%%%%%%%%%%%%%%%%%%%%%%%%%%%%%%%%%%%%%

\smallskip
\bibliographystyle{IEEEtran}
\bibliography{Analytical-Module-Journal}

%%%%%%%%%%%%%%%%%%%%%%%%%%%%%%%%%%%%%%%%%%%%%%%%%%%%%%%%%%%%%%%%%%%%%%%%%%%%%%%%%%%%%%%%%%%%%%%%%%%%%%%%%%%%%%%%%%%%%%%%%%%%%%%%%%%

% Not allocated / safety margin: 0.5 page

%%%%%%%%%%%%%%%%%%%%%%%%%%%%%%%%%%%%%%%%%%%%%%%%%%%%%%%%%%%%%%%%%%%%%%%%%%%%%%%%%%%%%%%%%%%%%%%%%%%%%%%%%%%%%%%%%%%%%%%%%%%%%%%%%%%
%%%%%%%%%%%%%%%%%%%%%%%%%%%%%%%%%%%%%%%%%%%%%%%%%%%%%%%%%%%%%%%%%%%%%%%%%%%%%%%%%%%%%%%%%%%%%%%%%%%%%%%%%%%%%%%%%%%%%%%%%%%%%%%%%%%

\end{document}